# CyberShake Earthquake Fault Rupture Modeling and Ground motion Simulations for the Southwest Iceland Transform Zone

Otilio Rojas[1], Marisol Monterrubio-Velasco[1], Juan E. Rodríguez[2], Scott Callaghan[3] Claudia Abril[1], Benedikt Holldorson[4,5], Milad Kowsari[4], Farnaz Bayat[4], Kim Olsen[6], Alice-AgnesGabriel[7,8], Josep de la Puente[1]

[1] Barcelona Supercomputing Center (BSC-CNS), Spain (otilio.rojas@bsc.es, marisol.monterrubio@bsc.es, claudia.abril@bsc.es)

[3] University of Southern California, USA (scottcal@usc.edu)

[4] University of Iceland, Iceland (milad@hi.is, fab14@hi.is)

[5] Icelandic Meteorological Office, Iceland (benedikt@vedur.is)

[6] San Diego State University, USA (kbolsen@sdsu.edu)

[7] Ludwig-Maximilians University, Germany (gabriel@geophysik.uni-muenchen.de)

[8] Scripps Institution of Oceanography, UC San Diego (algabriel@ucsd.edu)

## Abstract

CyberShake is a high-performance computing workflow for Probabilistic Seismic Hazard Assessment (PSHA) developed by the Statewide California Earthquake Center (SCEC). Here, we have implemented CyberShake to generate a set of 2103 fault ruptures and simulated the corresponding two horizontal components of synthetic seismic ground motion time histories on a 5-km grid of 625 stations in Southwest Iceland. The ruptures were defined on a new synthetic time-independent 500-year catalog consisting of 223 hypothetical finite-fault sources of $M_w$5-7, generated using a new physics-based bookshelf fault system model in the Southwest Iceland transform zone. This fault system model and its realizations of finite-fault catalogs effectively enable the CyberShake time-independent physics-based approach to PSHA in the region. The study aims to migrate CyberShake from its original study region in California to Southern Iceland and validate its kinematic-fault rupture, anelastic wave propagation and ground motion simulations. Toward this goal, we use CyberShake to generate multiple finite-fault rupture variations for each hypothetical fault. CyberShake exploits seismic reciprocity for wave propagation by computing Strain Green Tensors along fault planes, which in turn are convolved with rupture models to generate synthetic seismograms. For each surface recording station, every adjoint simulation uses a 0-1 Hz Gaussian point source polarized along one horizontal grid direction. Comparison of the results in the form of rotation-invariant synthetic pseudo-acceleration spectral response values at 2, 3. and 5 sec periods are in very good agreement with the Icelandic strong-motion dataset and a suite of new empirical Bayesian ground motion prediction equations (GMPEs). The vast majority of the CyberShake results fall within one standard deviation of the mean GMPE predictions, previously estimated for the area. Importantly, at large magnitudes for which no data exists in Iceland, the CyberShake synthetic dataset may play an important role in constraining the GMPEs for future applications. Our results comprise the first step towards comprehensive and physics-based PSHA for Southwest Iceland.

## Key points

1) We apply the SCEC CyberShake Platform toward physics-based PSHA for Southwest Iceland transform zone
2) The CyberShake ground motion simulations are consistent with a suite of new GMPEs for the region
3) At large magnitudes the CyberShake results may help constrain GMPEs for future applications

# Introduction

Iceland is the most seismically active region in northern Europe. It is a volcanic island in the North Atlantic Ocean and is the areal part of a massive flood-basalt plain produced by the Icelandic Hot Spot, a vertical magmatic thermal anomaly under central Iceland (Tryggvason et al. 1983; Thordarson and Hoskuldsson 2002). There it coincides with the Mid-Atlantic Ridge (MAR), the extensional plate boundary between the North American and Eurasian plates. The interplay of the MAR and the Hot Spot drives the volcanic and seismic activity of Iceland (Figure 1). Spatially, its seismic activity follows the present-day axis of active tectonic extension of the MAR in Iceland: it approaches from southwest where the subaerial Reykjanes ridge (RR) becomes aerial and joins the Reykjanes Peninsula Oblique Rift (RPOR), continues East through the South Iceland Seismic Zone (SISZ), turns north along the Eastern Volcanic Zone (EVZ), Northern Volcanic Zone (NVZ) and back towards West via the complex and largely subaerial Tjörnes Fracture Zone (TFZ). On the western edge of the TFZ, the MAR continues north of Central Iceland.

Due to the eastward ridge-jump of the extensional plate boundary on land in Iceland, two large transform zones have been formed: The complex and extensive Tjörnes Fracture Zone mostly offshore North Iceland, and the Southwest Iceland transform zone, comprising the SISZ and the RPOR. Historically, the largest earthquakes in Iceland of $M_\text{w}$ ~ 7 have repeatedly taken place in these transform zones according to historical annals of ~1000 years, teleseismic recordings of ~100 years, and the last ~30 years of local recordings of seismic strong-motions (Tryggvason et al. 1958; Einarsson 1991, 2008, 2014; Stefansson et al. 2008; Sigbjörnsson et al. 2014; Steigerwald et al. 2020; Einarsson et al. 2020; Jónasson et al. 2021). As a result, the SISZ-RPOR is one of two regions in Iceland where the seismic hazard is highest. In addition, the SISZ-RPOR is completely aerial and much more densely populated than the TFZ. Despite the RPOR mostly being a barren elevated plateau of basaltic Holocene lava shields, it contains several coastal towns most notably of which is the capital region of Reykjavik where ~2/3$^\text{rd}$ of the national population lives. Further East and collocated with the SISZ lies the South Iceland lowland, the country's largest agricultural region. The SISZ-RPOR contains typical infrastructure pertinent to modern society and its lifelines such as electric power transmission lines, hydroelectric and geothermal power plants, dams, bridges, roads and pipelines.

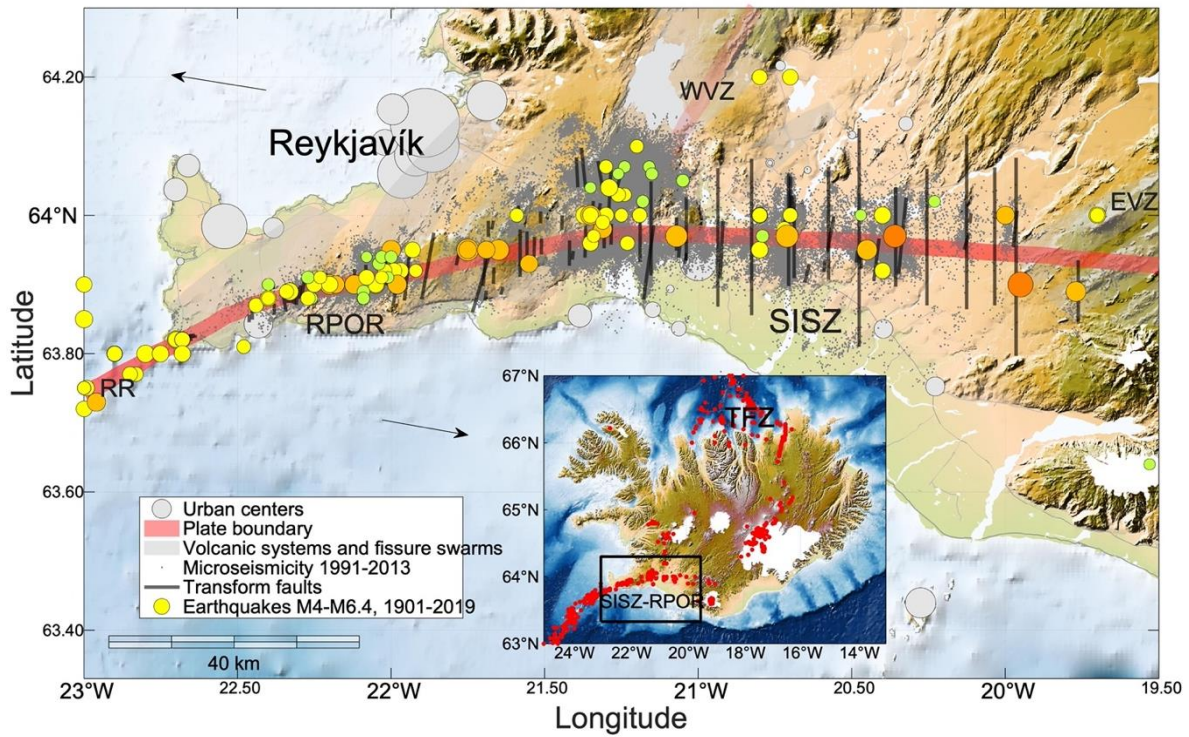

Figure 1. The Southwest Iceland transform zone, consisting of the South Iceland Seismic Zone (SISZ) in the South Iceland lowlands and the Reykjanes Peninsula oblique rift zone (RPOR). The red line denotes the centerline of the plate boundary across southwest Iceland with vectors denoting the approximate direction of transcurrent motion across the SISZ-RPOR. The instrumental microseismic catalogue from 1991-2013 is shown as dots, and significant earthquakes from 1904-2019 of the ICEL-NMAR catalogue as circles, spanning magnitudes from Mw 4 (green circles) to Mw 7 (dark orange circles). The black lines denote the mapped and inferred locations of dextral strike-slip faults of the bookshelf fault system. Distinct volcanic systems in volcanic zones (see acronyms in text) are oriented northeast-southwest and shown as gray shaded regions.

A reliable estimation of seismic risk benefits from a reliable probabilistic seismic hazard assessment (PSHA). The current seismic hazard map for Southwest Iceland is over two decades old, but more importantly, it needs a comprehensive revision as several recent research findings have shown (for an overview see e.g., Kowsari et al. 2021, 2022). Namely, it has conclusively been shown that the fault system in Southern Iceland, previously thought to be confined to the SISZ region, it actually continues towards west and all along the RPOR (Steigerwald et al. 2020; Einarsson et al. 2020). In addition, the maximum seismogenic potential of SISZ-RPOR, and in particular in the RPOR, is seen to decrease systematically from east to west (see in Bayat et al. 2022, 2024, and references therein). That necessitates a subdivision of the Southwest Iceland transform zone into subzones, each of which having a specific maximum earthquake magnitude and a specific earthquake magnitude-frequency distribution (MFD). However, the most recent compilation of a harmonized instrumental earthquake catalogue of significant earthquakes in Iceland since 1900 (Jónasson et al. 2021) is too sparse, in particular for the RPOR, to allow the derivation of reliable subzone MFDs. This is of practical importance as the Central and Eastern RPOR fault system lies near the capital region of Iceland.

Fortunately, the above limitations have effectively been addressed through the development of a new physics-based earthquake fault system model of the Southwest Iceland transform zone. It has been calibrated to the rate of tectonic transform motion across the zone and constrained by the geometry of the fault system and the variation of seismogenic depth along the transform zone (Bayat et al. 2022). The model allows random realizations of the fault system configuration where each causative fault is completely specified by its dimensions, maximum potential seismic magnitude, and long-term slip rate, along with uncertainty estimates. It also allows the derivation of subzone specific MFDs, the cumulative seismic activity of which is not only consistent with, but effectively explains the earthquake catalogue of the Southwest Iceland transform zone (Bayat et al. 2024). In particular, the ability of the model to generate realizations of long-term finite-fault earthquake catalogues (Kowsari et al. 2022b) means that it lends itself particularly useful for a physics-based Monte Carlo stochastic approach to PSHA, i.e., earthquake rupture modeling and seismic ground motion simulation using the kinematic modeling approach. Not only is a Monte Carlo approach to PSHA a more physically realistic approach but it also enables better incorporation of uncertainties into the PSHA estimate. The underlying hypothesis is to assume that the seismogenesis of each fault follows a certain probabilistic model of seismicity (e.g., Poissonian, Markovian), and whose parameterization depends on pre-existing data and models. The Monte Carlo approach to PSHA has been used since 1995 in Iceland (Sigbjörnsson et al. 1995; Solnes et al. 2004) but more recently in various regions of the world for which various seismicity models apply (Atkinson 2012; Bayraktar et al. 2017; Karaca 2021).

A physics-based approach to PSHA is computationally very intensive. In the last 1-2 decades, however, computational seismology using high-performance computing (HPC) has explored physical models to explain the mechanisms that govern earthquake rupture and seismic wave propagation. When modeling earthquake ruptures, kinematic models prescribe finite slip evolution in space and time and allow to explore variability in seismic moment, fault orientation, hypocenter location, source time function, and rupture path, while dynamic rupture models (e.g., Li et al. 2023) use the physics of frictional rock failure to model how earthquakes start, dynamically propagate and arrest and may probe in addition to the variability in kinematic models also differences in fault frictional parameters, coseismic multi-fault interaction and local stress distributions. The sampling of source parameters and their variability combined with the simulation of wave propagation accounting for wave-physics effects has allowed a physics-based approach to PSHA (Graves et al. 2011; Bradley et al. 2018; Callaghan et al.

2021; Milner et al. 2021). The computational platform for such a calculation needs to efficiently simulate the ground motions at each site for a large (at least in the order of thousands) ensemble of rupture variations. For instance, Graves et al. (2011) and Callaghan et al. (2021) assembled an Earthquake Rupture Forecast (ERF) by compiling the likely source variability on California fault systems, consistent with slip inversions of past earthquakes (Field et al. 2009). Alternatively, Milner et al. (2021) used the Rate-State earthquake simulator (RSQSim) to simulate hundreds of thousands of years of synthetic seismicity also in California. This source variability allows complementing earthquake catalogs with synthetic data of moderate- and large-magnitude ranges, enriching PSHA results.

CyberShake is a physics-based PSHA tool that integrates the Graves-Pitarka (GP) pseudo-dynamic fault-rupture generator with an anelastic wave propagation solver for ground-motion simulation. CyberShake was originally developed to undertake PSHA studies in California (Graves et al. 2011; Silva et al. 2016; Jordan and Callaghan 2018), but it has the potential to be implemented in other tectonic areas. This computational workflow has benefitted from a significant evolution in HPC facilities for efficient simulation of hundreds of thousands of events as required for a Monte Carlo approach to PSHA. Earthquake modeling in CyberShake is implemented in reciprocal mode. For every hazard location, two simulations are performed, each one using a source delta function polarized along a horizontal direction to compute the response of the strain tensor at each fault surface point (Strain Green Tensors - SGT). SGT time histories require a significant disk storage amount, in the order of 1-2 terabytes per hazard site (Maechling et al. 2007; Callaghan et al. 2015), according to the total number of fault segments on the ERF database. The convolution of these SGTs with numerous GP rupture models produces particle-velocity seismograms at each hazard site (Maechling et al. 2007; Callaghan et al. 2008, 2014). CyberShake can produce hundreds of thousands of synthetic seismograms at sites spread over Central and Northern California (Callaghan et al. 2015, 336 sites; Callaghan et al. 2017, 438 sites; Milner et al. 2018, 869 sites). More recent applications have been focused on the hybridization of CS results by including higher-frequency stochastic ground-motions with application primarily in dynamic analysis of structural and geotechnical systems (Callaghan et al. 2020).

This study constitutes one of the first applications of the CyberShake platform outside California. We note that Bradley et al., 2018, employed a different computational framework using forward simulations for physics-based PSHA in New Zealand. In the following, we first describe the tectonic setting and seismological characteristics of the Southwest Iceland transform zone. We then present details of the migration of CyberShake to the regional earthquake source scaling in Southwest Iceland, in particular the calibration of the rupture generator to local earthquake magnitude-area scaling law and crustal velocity and density models scaling. We apply the new physics-based fault system model of the SISZ-RPOR transform zone to generate a finite-fault realization of the fault system equivalent to a 500 year-long synthetic earthquake catalogue, covering earthquake moment magnitudes, in the range of Mw 5 to Mw 7. We then apply the CyberShake earthquake rupture generator in the simulation of multiple rupture variations of each earthquake in the finite-fault catalogue, and in the kinematic simulation of the seismic ground motions on a rectangular grid of 625 hypothetical stations in Southwest Iceland. This simulation employs a 5 km interstation spacing. Thus, for each earthquake fault rupture variation and each hypothetical station, two horizontal-component synthetic seismic time histories are simulated, each one using a 0-1 Hz point source polarized along one of the two horizontal grid directions. Key strong-motion intensity measures are then derived from the time histories. In particular, the peak ground velocity and pseudo-acceleration spectral response (PSA) estimates of single-degree-of-freedom oscillators of natural periods from 2 to 10 s. We validate the results by comparing the PSA with new magnitude-distance dependent empirical ground motion prediction equations (GMPE) derived from regional strong-motion data. Finally, we examine the fundamental characteristics of the synthetic motions in the near-fault region.

## The Southwest Iceland transform zone

**Tectonics of the bookshelf fault system**

The transcurrent tectonic plate motion across the Southwest Iceland transform zone is oriented ~ East-West along the RPOR and SISZ. However, instead of a large East-West aligned sinistral strike-slip fault, the release of tectonic strain is accommodated on short near-vertical dextral strike-slip faults striking North-South, perpendicular to the transcurrent axis (see Figure 1). This has been conclusively confirmed in past studies that have mapped old surface fault traces, relative relocations of localized seismic swarms, inferred fault locations of large historical earthquakes, along with macroseismic felt area extent and orientation from historical accounts, and detailed source and strong-motion studies of recent Mw 6.3-6.5 earthquakes in 2000 and 2008 (e.g., Einarsson 1991, 2010; Morgan and Kleinrock 1991; Stefánsson et al. 1993; Pagli et al. 2003; Clifton et al. 2003; Pedersen et al. 2003; Roth 2004; Árnadóttir 2004; Dubois et al. 2008; Hjaltadóttir 2009; Hreinsdóttir et al. 2009; Halldorsson and Sigbjörnsson 2009; Decriem et al. 2010; Steigerwald et al. 2020; Einarsson et al. 2020).

The average distance between the faults appears to be approximately 5 km (Stefánsson et al. 2006), but there is evidence that these inter-fault distances decrease toward the west in the RPOR, even down to a few hundred meters (Steigerwald et al. 2020). Hypocentral distributions of small earthquakes in the SISZ and the RPOR are mostly at 1-5 km depth in the western part of the RPOR but the seismogenic depth increases gradually towards the east, culminating at a maximum 12-15 km depth in the easternmost part of the SISZ (Stefánsson et al. 1993; Einarsson 2015; Panzera et al. 2016; Steigerwald et al. 2020). As a result, the seismogenic potential of the zone increases toward the east, in agreement with the historical earthquake catalog (Einarsson et al. 1981; Dubois et al. 2008; Panzera et al. 2016).

All strong historical earthquakes in Southwest Iceland that have caused damage are considered to have occurred on the bookshelf fault system, with volcanic earthquakes in the RPOR being smaller and thus contributing little to the PSHA in the region. However, reliable earthquake catalogues for subzones of different maximum potential earthquake magnitudes are relatively sparse, in particular along the RPOR. A new physics-based model of the bookshelf earthquake fault system in the Southwest Iceland transform zone has been developed (Bayat et al. 2022). It has been calibrated to the rate of tectonic transform motion across the zone and constrained by the geometry of the fault system and the variation of seismogenic potential along the transform zone. The model allows random realizations of the fault system configuration where distance between faults is varied, with the model specifying each fault dimensions, maximum potential seismic magnitude, and long-term slip and moment rate, along with parametric uncertainty estimates. The model allows the derivation of subzone-specific MFDs, the cumulative seismic activity of which is not only consistent with, but effectively explains the long-term earthquake catalogue of the Southwest Iceland transform zone (Bayat et al. 2024). The model has recently been applied to generate realizations of long-term finite-fault earthquake catalogues and used for point-estimates of PSHA at representative far-field and near-fault sites, albeit using conventional methods based on empirical GMPEs (Kowsari et al. 2022a). However, finite-fault earthquake catalogues are much more useful for a physics-based approach to PSHA, through earthquake rupture modeling and seismic ground motion simulations, using either the dynamic or kinematic modeling approach.

**Methodology and localization**

Earthquake physics-based simulations require adequate inputs to provide reliable outcomes. The fundamental parameters are related to the description of the (an)elastic physical properties of the subsurface and the detailed description of space-time energy sources, typically slip and slip rate on a fault. The specific description of such parameters (e.g., discretization, smoothness) is related to the numerical method used for the simulations and its implementation. Moreover, when assessing the expected ground motion for more than a single specific earthquake, we need to populate a synthetic catalogue, characterized by locations, magnitudes and fault mechanisms. Lastly, a series of sites of interest, which may or not coincide with existing seismic stations, is needed to obtain the synthetics records for further analysis. The whole workflow then needs to be orchestrated, typically on a high-performance computing system, The details of the methods and processes involved are presented in the following.

**Pseudo-dynamic fault-rupture model**

The kinematic rupture description of an earthquake source prescribes the spatial and temporal evolution of the slip vector across the fault surface. A pseudo-dynamic kinematic rupture description is expected to abide to well-known principles of seismic fault dynamics to constrain, e.g., source-time functions, rise time or slip rate (Thingbaijam et al. 2017). The Graves and Pitarka (GP, Graves and Pitarka 2010, and 2016) rupture generators fall into this category while being able to produce full kinematic models including, e.g., the dependence of rupture velocity and slip rate with depth and abiding to a prescribed magnitude, fault orientation, area and depth. Regarding the asperity scaling law, GP methods depend on Von Karmann filters applied to initial random slip distributions. For the purposes of the current study, we adopt the filter coefficient values proposed by Mai and Beroza (2002), namely $a_S$ = 2.5 (strike direction) and $a_P$ = 1.5 (dip direction). In this study, we use the current and specific CyberShake implementation of the Graves and Pitarka (2016) method for rupture model generation.

**Three-dimensional wave-propagation simulation solver**

The wave-propagation solver available in CyberShake is the finite-difference velocity-stress staggered-grid Anelastic Wave Propagation (AWP-ODC) code, which presents a fourth-order spatial accuracy coupled to a second-order leapfrog time-marching scheme. This code has been highly optimized to perform site-to-fault reciprocal simulations for computing the Strain Green Tensors along all fault planes (Jordan and Callaghan 2018). Alternative code versions perform standard forward fault-to-site earthquake simulations (Cui et al. 2010), which have been also efficiently implemented in parallel architectures. The main AWP-ODC limitation of modeling flat topographies has been recently overcome by discretizing this interface using curvilinear coordinates (O'Reilly et al. 2022). In the present study, we limit our scope to flat topography. We use standalone fault-to-site configurations for single-event validation studies and CyberShake-integrated site-to-fault configurations for massive multi-event cases. No numerical difference is expected between the two configurations. For all cases, we have used 9 points per minimum S-wavelength that allows highly accurate results up to our prescribed cutoff frequency of 1 Hz.

**Implementation details**

The installation of the CyberShake platform at the Marenostrum 4 Supercomputer is the result of a collaboration between the Statewide California Earthquake Center (SCEC) and the Barcelona Supercomputing Center (BSC--CNS, Centro Nacional de Supercomputación) and other partners in the framework of the ChEESE Center of Excellence (Folch et al., 2023). In order to run Cybershake in Marenostrum, the UnifiedCSWflow open-source workflow manager was developed (Rodriguez et al. 2021). It organizes the end-to-end execution process according to data dependencies of the CyberShake workflow, and effectively replaces the original workflow orchestration based on Pegasus and HTCondor for the current application. UnifiedCSWflow handles the CyberShake relational database that stores the input Earthquake Rupture Forecast (ERF) data, the identification metadata of each processing step and the large amount of output Strain Green Tensors (SGTs) and hazard intensities.

**Velocity and density model**

A fundamental requirement for earthquake physics-based seismic simulations is an accurate model of the properties of the subsurface. Specifically, we need a representation of the compressional (P) and shear (S) wave speeds, material density and attenuation. Our study area comprises both the RPOR and SISZ regions. For each region, we used the one-dimensional P- and S-wave velocity models ($\alpha$ and $\beta$, respectively) implemented for location of the SIL catalog earthquakes, as seen in Figure 2. We can observe the properties of the upper crust, between 3 and 4 km thick, followed by a low velocity-gradient lower crust. The lower crust limit is apparent at 16 to 22 km in this model. For the purpose of seismic simulations, the models need to be coupled and smoothed. For the first task, we have included a 10 km wide crossover zone starting at longitude -21.38º toward the west, where $\alpha$ and $\beta$ are linearly harmonic averaged to avoid undesired numerical artifacts. For the resulting velocity model, density is estimated following the empirical law for Icelandic rocks by (Darbyshire et al. 2000). Along depth, we have applied Gaussian smoothing to the three elastic properties, where P- and S-wave slownesses are smoothed instead of the velocity to preserve travel time. For our South Iceland study, we use visco-elastic attenuation parameters employed in CyberShake simulations in California, namely, $Q_S = 0.05\beta$ ($\beta$ in m/s) and $Q_P = 2\ Q_S$.

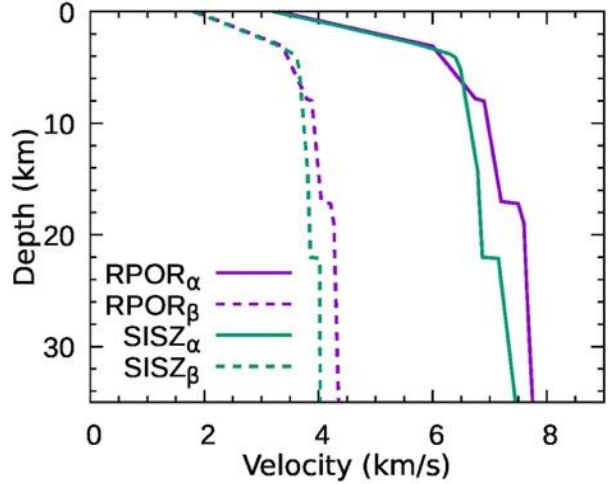

Figure 2. 1D velocity models for the RPOR and the SISZ regions used at the SIL catalog for earthquake location. Synthetic finite-fault earthquake catalogue for South Iceland

In this study, we use a synthetic finite-fault earthquake catalogue of 500-year duration, generated as a realization of the 3D fault system model of the SISZ and RPOR, based on the work of Bayat et al. (2022, 2024) and Kowsari et al. (2022a). The realization simulates the spatial variation of seismogenic potential along the zone, i.e., increasing maximum earthquake magnitude from Mw ~ 5.5 in the western part of the RPOR to Mw ~ 7 in the eastern part of the SISZ (Bayat et al. 2022). Moreover, the synthetic catalogue allows the systematic spatial variation of fault slip-rates to be modeled by discrete subzonation of the fault system, and the equivalent parameters of earthquake MFDs, something that the limited earthquake catalogues cannot produce (Bayat et al. 2024). These zone-specific MFDs are inherent in the realization and are compatible with the physical constraints of the fault system and the long-term earthquake catalogues for the region, allowing exploring the uncertainties in the parametrization.

The synthetic earthquake catalogue contains earthquakes in the moment magnitude range of Mw ~5 to Mw ~7 i.e., the earthquake magnitudes that contribute the most to PSHA and up to the maximum earthquake potential of the region, specifically the easternmost part of the SISZ (Kowsari et al. 2022a). The uncertainty of fault locations along the zone is sampled in the synthetic catalogue by selecting random locations within each zone around the centerline of the tectonic plate margin, constrained by the locations of the historic and instrumental seismicity and extent of mapped and inferred faults in the region. Figure 3 shows the rupture extents from the finite-fault earthquake catalogue used in this work that comprises 223 parallel north-south, near vertical dextral strike-slip faults. Thus, this synthetic finite-fault earthquake catalogue facilitates physics-based simulations of the low-frequency seismic motions in particular, in the near-fault region.

Notably, the earthquake magnitude-area scaling in the Southwest Iceland transform zone does not follow the most common scaling laws of shallow crustal interplate strike-slip earthquakes (e.g., those in Wells and Coppersmith 1994). The estimated fault extents of historical earthquakes, and in particular those of the recent $M_w$ 6.3-6.5 earthquakes in 2000 and 2008, show a much smaller relative fault area and relatively large slip than expected by most scaling laws (see e.g., Pedersen et al. 2003; Dubois et al. 2008; Hreinsdóttir et al. 2009; Decriem et al. 2010). However, the "effective source area" scaling law of shallow crustal interplate strike-slip earthquakes of Mai and Beroza (2000) has been found to describe well the *total area* of earthquakes in the SISZ-RPOR and has been applied here in accordance with the original study (Bayat et al. 2022) and subsequent applications (Kowsari et al. 2022a; Li et al. 2023).

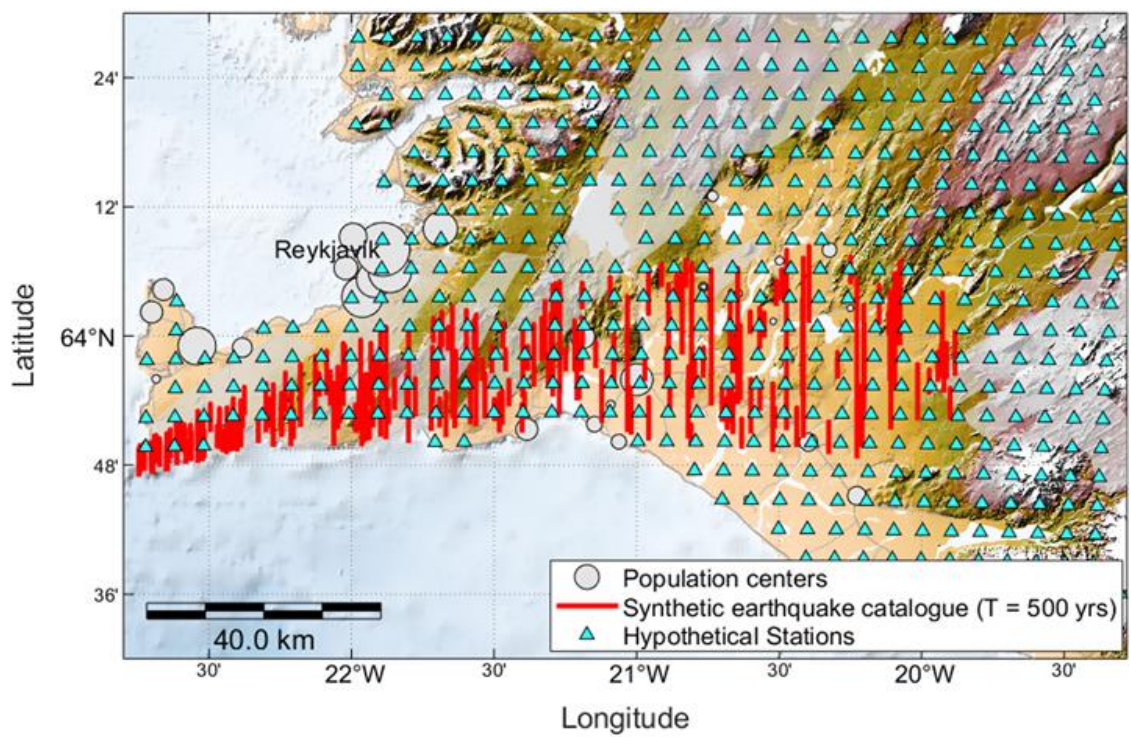

Figure 3. CyberShake simulation domain in Southwest Iceland. The hypothetical station grid consists of 625 sites and the realization of the RPOR-SISZ fault system model used in the simulations comprises 223 North-South striking dextral vertical strike slip fault planes (red lines).

We show in Figure 4 examples of two rupture variations for hypothetical $M_w$6 and Mw 7 events. They are defined by the specific faults of the synthetic catalogue that they correspond to, a realization of slip distributions and hypocentral location on the fault plane. They reflect the expected characteristics of earthquake source scaling in the region, in accordance with slip inversion studies of recent strong earthquakes in Southwest Iceland, which are characterized by relatively small total rupture area and large slip for the given magnitudes, and 2-4 large slip-patches that drive the near-fault wavefield amplitudes.

**Results**

We run CyberShake using a couple of 0-1 Hz Gaussian point sources as required by the adjoint wave-propagation modeling and the laterally-smoothed 1d velocity and density model that combines the SISZ and RPOR models illustrated in Figure 2. A total of 2103 rupture variations are modeled each with two horizontal components of seismic ground motion time histories per site. We compare the simulated ground motion intensity measures to a set of empirical GMPEs recently developed for this region. As the GMPEs use a rotational invariant measure of horizontal gr(Rupakhety and Sigbjörnsson13), we calculate the same measure from the synthetic ground motions at each station. The comparison of the CyberShake synthetic strong-motio, dataset to the GMPEs Figure 55. The results have been plotted in terms of the two key independent parameters of the empirical GMPEs, moment magnitude $R_{JB}$ and = 0.1 km as the shortest horizontal distance from a site to the vertical surface projection of the fault. The comparison is made in terms of pseudo-acceleration spectral response PSA) at oscillator periods of 2 to 5 seconds, and at four magnitude bins/ranges, Mw 5.1-5.4, Mw 6, Mw 6.3-6.5 and Mw 7. These magnitudes reflect mainshock seismic ground motion recordings in the Icelandic strong-motion dataset for Southwest Iceland, shown as red dots. In contrast, the parameters of the CyberShake synthetic dataset are shown as gray dots, i.e., the discrete values of PSA at each station correspond to a dot in the far-right column of Figure 6 at those periods and distances. The comparison brings to light the small number of actual earthquake ma$M_w$, and the fact that no data exists in Iceland for earthquakes larger than 6.5. The empirical GMPEs have been developed based on the local dataset, but these limitations have been alleviated somewhat through Bayesian inference using informed priors of magnitude-distance scaling from other GMPEs calibrated to strong-motion data from shallow crustal earthquakes in other interplate regions. The mean predictions of the GMPEs are shown for comparison $\pm 1\sigma$, along with their (for details, see Kowsari et al. 2020). We note that multiple same color GMPE attenuation curves cover the denoted magnitude ranges in steps of 0.1 magnitude units.

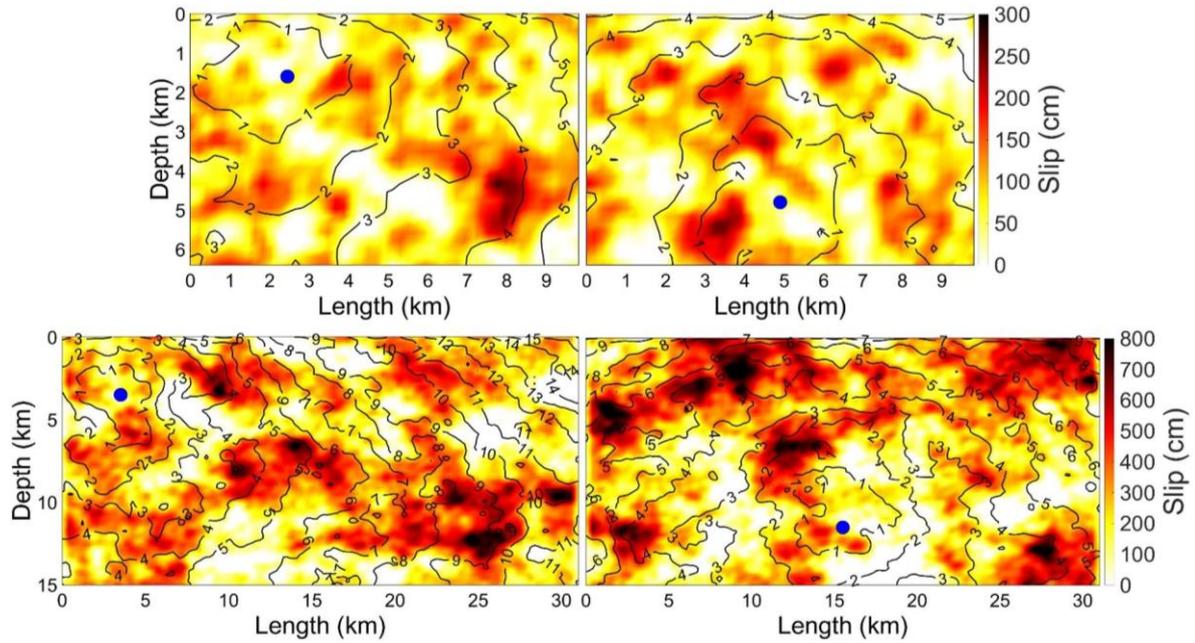

Figure 4. Two examples of rupture variations of $M_w$ 6 (top) and $M_w$ 7 (bottom) earthquakes from the synthetic finite-fault earthquake catalog. Each rupture variation is defined by different realizations of random slip distributions and different hypocentral location on the fault plane. The examples shown correspond to a unilateral (towards North, left) and bilateral fault rupture (right), respectively.

The recorded data and the CyberShake dataset show a very good agreement in the magnitude range $M_w$ 5-6.5. The same is the case for the comparison against the mean GMPE predictions. The vast majority of the CyberShake results fall into a $\pm 1$ standard-deviation band around the GMPE mean value predictions, which statistically validates its migration to the SISZ-RPOR region. Of interest is the comparison at $M_w$ 7 where the GMPEs appear to underpredict the PSA synthetic data 2 s and 3 s period over the entire distance range. However, at longer periods, the agreement is better overall, although the attenuation of the long-period response spectral values appears to be larger with distance than that of the GMPEs. Since the GMPEs are not constrained by local data at these large magnitudes, the CyberShake results may play a role in constraining them for future applications. We note that not all of the new GMPEs were calibrated to such long periods and as a result, fewer GMPEs are compared with the data at those periods.

We note that the observed PSA amplitudes in the extreme near-fault region of the $M_w$ 6.3-6.5 earthquakes ($R_{JB} \sim$ 1-3 km) appear to significantly underpredict the values of the GMPEs at periods above T=2 s. Namely, the energy of the near-fault data is driven by the near-fault velocity pulses that have a fundamental pulse-period of T~1.5-2 s and moreover have a narrow-band pulse character i.e., rather devoid of energy at longer periods (Halldórsson et al. 2007).

The pulse-period is controlled primarily by the characteristic dimension of the main slip patch on the fault plane i.e., the one that generates the velocity pulse (Mavroeidis and Papageorgiou 2003; Mavroeidis et al. 2004). The narrow-band nature of the Icelandic near-fault pulses thus implies a dominant characteristic slip-patch dimension on the causative fault plane. Indeed this is the case from multiple studies that inferred the static slip distributions from ground motion and deformation data (see Pedersen et al. 2003; Dubois et al. 2008; Hreinsdóttir et al. 2009; Decriem et al. 2010). The synthetic (CyberShake) PSA amplitudes for $M_w$ 6.0-7.0 earthquakes, however, do not show such characteristics, which by analogy implies that their energy content is broader, i.e., at longer periods as well i.e., the near-fault data is generated by slip patches of greater variability in their dimensions (which is in fact observed in the examples in Figure 4). This is a positive feature of the CyberShake slip distributions and we note that the built-in randomization features of the source specification (e.g., subfault strike, dip and rake are allowed to vary within a range) may have weakening effects on ground motion coherence at near-fault locations.

Inspecting further the rupture scenario shown in Figure 4 i.e., the Mw7 bilateral rupture, Figure 6 shows the location of the synthetic earthquake fault along with its random slip distribution and the subfault rupture onset time.

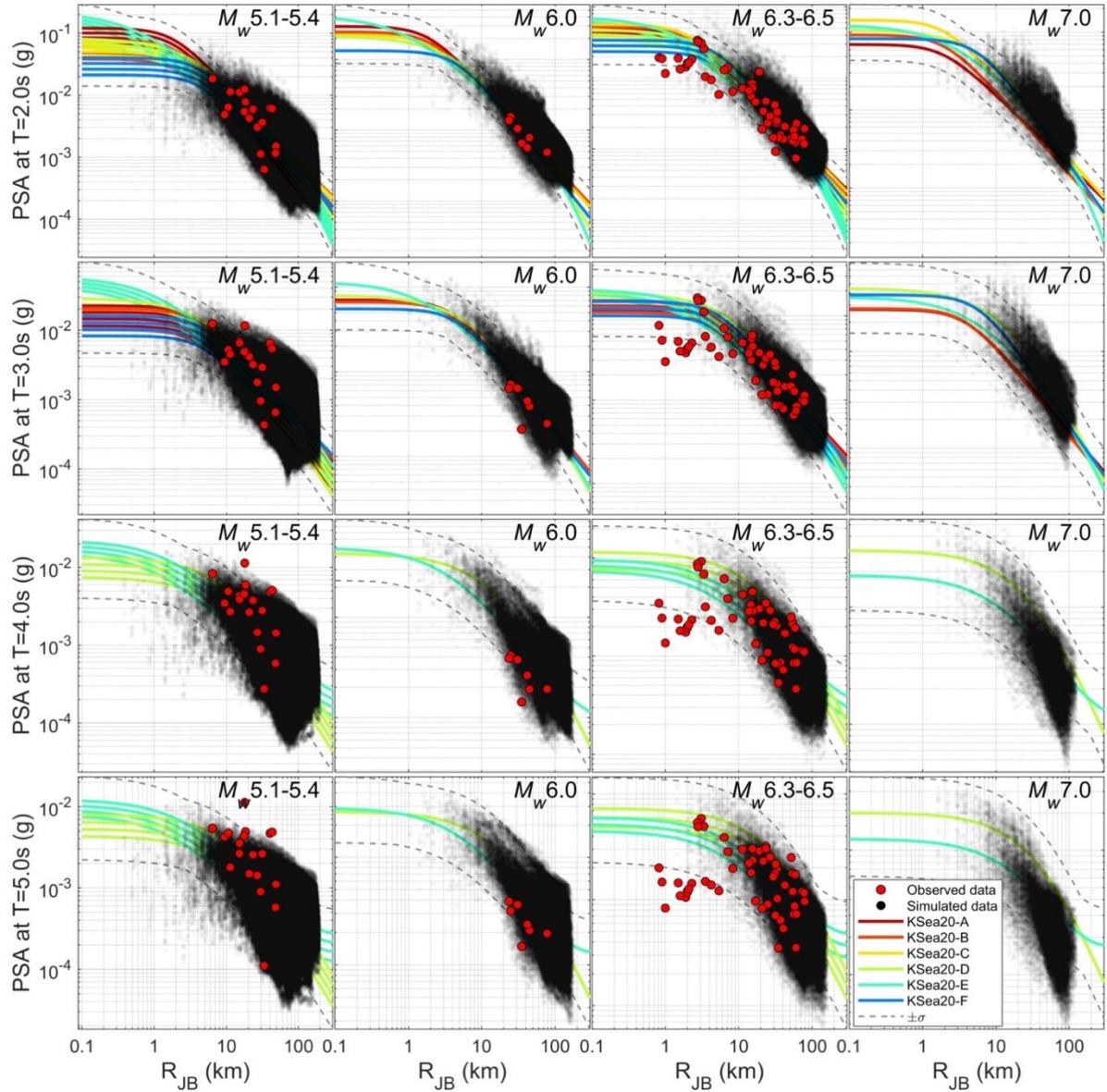

Figure 5. CyberShake ground motion scaling with magnitude and fault distance for the Southwest Iceland transform zone. The synthetic pseudo-acceleration spectral response (PSA, black circles) values at four discrete periods of oscillation ($T$ = 2, 3, 4 and 5 s, top to bottom rows) and at four magnitude ranges ($M_w$ 5.1-5.4, 6, 6.3-6.5, and 7, left to right columns) are compared with the actual values from the strong-motion dataset for Southwest Iceland (red circles) and the six new Bayesian empirical ground motion prediction equations (KSea20, Kowsari et al. 2020). The ground motion parameter is the rotation invariant measure of two horizontal PSA estimates. Multiple same color GMPE attenuation curves cover the denoted magnitude ranges in steps of 0.1 magnitude units.

The corresponding rotation invariant ground motion measure of PSA at 2, 3 and 5 s oscillator periods are shown in Figure 7 where dots indicate the locations of the stations. The synthetics show that the greatest amount of energy is at T = 2 s period, progressively decreasing with increasing period. The characteristics of the amplitude distribution around the fault show a concentration of large amplitudes in the extreme near-fault region of the southern end of the bilateral rupture. This is consistent with rupture directivity and the observation that most of the fault slip on that particular realization is located in the shallow and southern part of the fault, with the rupture propagating primarily upwards. The large amplitudes at the stations next to the southern part of the fault are to different degrees dominated by the three slip patches that appear at shallow depth. Scrutiny of the synthetic time histories reveals that the fault rupture generates large-amplitude and long-period near-fault velocity pulses of periods from 2-4 seconds in this region. Overall, the synthetic on near-fault wavefield exhibits greater complexity than has been observed in admittedly the very limited near-fault data recorded in Iceland. That is to be expected as the near-fault motions of the CyberShake dataset reflect its sampling of source variability, which is promising for the future analyses and applications of this dataset comprising 2103 fault rupture simulations.

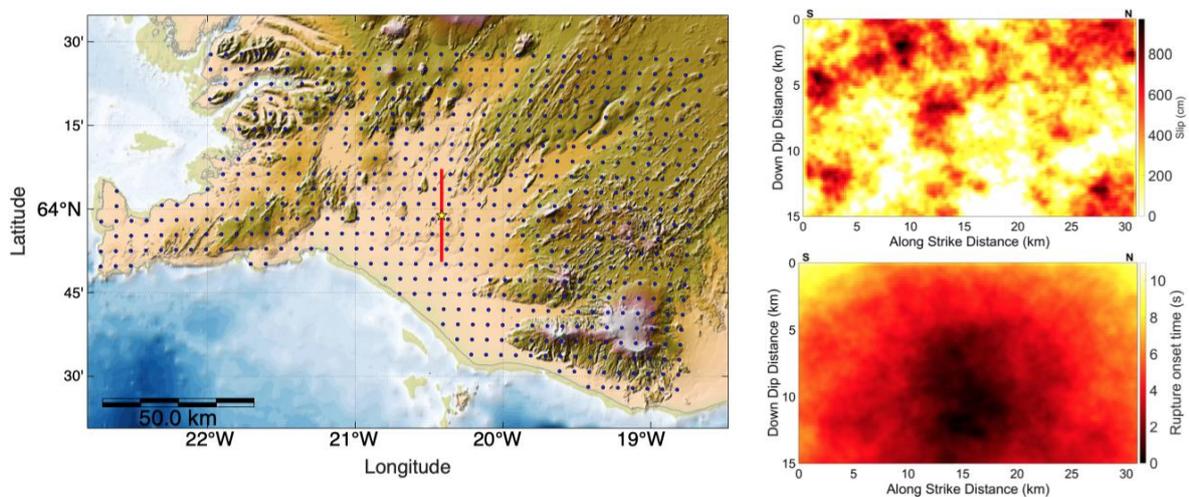

Figure 6. A hypothetical Mw 7 bilateral earthquake rupture scenario in the SISZ along with the synthetic slip distribution and propagating rupture front.

## Discussion and Conclusions

The migration of CyberShake (CS) to Southwest Iceland comprises two main porting steps. From a seismological standpoint, an earthquake rupture forecast (ERF) has been assembled using a synthetic finite-fault and time-independent earthquake catalogue corresponding to strong earthquake ruptures on the unique bookshelf fault system of the Southwest Iceland transform zone. The transform zone consists of the South Iceland Seismic Zone (SISZ) and the Reykjanes Peninsula Oblique Rift (RPOR) zone, and the catalogue equivalent to 500 years of seismicity contains 223 faults of magnitudes between Mw 5 and Mw 7. The earthquake magnitude-frequency distribution is controlled by the six distinct subzones of the SISZ-RPOR bookshelf fault system, each of which has a specific distribution and a maximum earthquake magnitude. The earthquake magnitude-area scaling law for the region has been embedded into the finite-fault catalogue. Moreover, slip inversions in SISZ-RPOR have revealed that faulting is mainly characterized by a few large asperities that drive strong near-fault ground shaking. These features of the applied slip distributions are controlled by the Von Karmann (VK) filter parameters, where we use the recommended values by Mai and Beroza (2002). Using these parameters, the method developed by (Graves and Pitarka 2016) is suitable for constructing kinematic ruptures on each fault of the synthetic finite-fault earthquake catalogue. Source variability is explored by varying hypocenter locations along the strike and dip directions in 4.5 km squared cells. CS simulations up to 0.5 Hz have been carried out using a velocity model that results from smoothly coupling the P- and S-wave profiles from two 1D layered models. A total of 2103 earthquakes are modeled, and the corresponding two horizontal component time histories of synthetic seismic ground motion simulated at 625 sites in Southwest Iceland.

From a technological standpoint, CS executions have been highly facilitated by the development of the open source UnifiedCSWflow workflow manager. UnifiedCSWflow handles the whole execution process according to data dependencies of processing steps and controls CS database access by storing the input ERF data, the identification metadata of each processing step and the large number of finally computed seismograms and hazard intensities. In this sense, UnifiedCSWflow has properly replaced the complex original workflow used in California to run CS.

We compare peak synthetic parameters of the CS dataset in the form of rotation invariant measures of pseudo-acceleration spectral response at 2, 3, 4 and 5 sec periods to a suite of new empirical GMPE models developed recently for the region (Kowsari et al. 2020). The synthetic magnitude-distance characteristics cover $M_\text{w}$5-7 and Joyner-Boore distances of less than 1 km to a maximum of ~160 km. The CS dataset is in very good agreement with the GMPEs, with the vast majority of CS results falling within a $\pm 1$ standard-deviation band around the GMPE mean predictions at most magnitudes. At the largest considered earthquake magnitudes, $M_\text{w}$7, the GMPEs appear to be slightly underpredicting the PSA values at 2-3 s oscillator periods, while the comparison at longer periods shows a better fit. However, as the GMPEs are not constrained by local data at these large magnitudes, the CyberShake synthetic dataset may play an important role in constraining the GMPEs in future applications. A preliminary look at near-fault amplitudes of selected scenarios shows great consistency with rupture variation characteristics i.e., the geometry of the station grid, the hypocenter location on the fault, and the location of the significant slip patches on the fault plane. The characteristics of the synthetic near-fault wavefield may indicate that it samples in greater detail the source complexity than perhaps has been observed in the very limited near-fault data recorded in Iceland to date. We conclude that the consistency of the CyberShake synthetic dataset of long-period motions with the GMPEs effectively and statistically validates its migration to the SISZ-RPOR region. This is a prerequisite for future work, towards increasing the duration of the synthetic earthquake catalogue and maximum frequency of the ground motions, so that a statistically consistent and physics-based PSHA study may be carried out in the SISZ-RPOR region, and to routinely apply Cybershake outside of California.

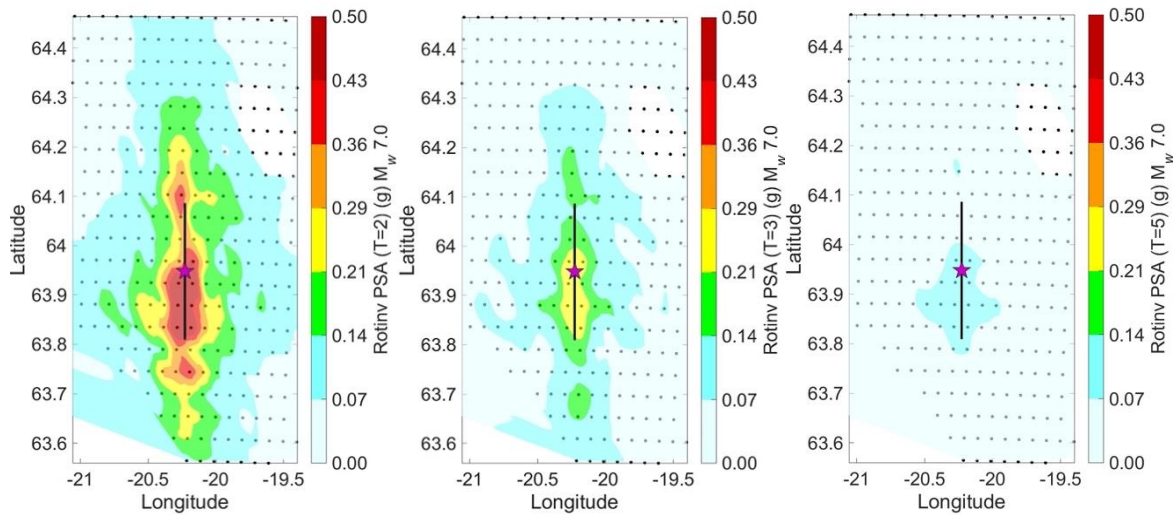

Figure 7. The rotation invariant ground motion measure of PSA at (left to right) 2, 3 and 5 s oscillator periods for the $M_\mathrm{w}$ 7 bilateral rupture scenario in Figure 6.

## Conflict of interest

The authors declare no conflict of interest.

## Acknowledgements

The research leading to these results has received funding primarily from the European Union's Horizon 2020 research and innovation programme under the ChEESE projects (Grant agreement No. 823844), the Horizon Europe projects DT-GEO (Grant agreement No. 101058129), Geo-Inquire (Grant agreement No. 101058518), and the Horizon 2021 EuroHPC JU-RIA grant (No. 101093038, ChEESE-2P). In addition, partial funding was received from the Icelandic Centre for Research (Rannis) Icelandic Research Fund Project Grant (No. 196089, SENSHAZ) and a Doctoral Grant (No. 239376), the Landsvirkjun Energy Research Fund (No. NÝR-04-2023), the Icelandic Road and Coastal Administration Research Fund (No. 1800-947), and the University of Iceland Research Fund (No. 92334). AAG acknowledges additional funding from the European Union's Horizon 2020 Research and Innovation Programme (TEAR ERC Starting, Grant 852992) the National Science Foundation (NSF Grant OAC-2311208 and EAR-2121568), and the National Aeronautics and Space Administration (80NSSC20K0495). Finally, we would like to thank to Philip J. Maechling for continuous communication and scientific feedback relevant to the achieved results of this work and acknowledge SCEC for providing additional technical software and scripts used for output data postprocessing.

**References**

Árnadóttir T (2004) Coseismic stress changes and crustal deformation on the Reykjanes Peninsula due to triggered earthquakes on 17 June 2000. Journal of Geophysical Research 109:. https://doi.org/10.1029/2004JB003130

Atkinson GM (2012) Integrating advances in ground-motion and seismic hazard analysis. In: Proceedings of the 15th World conference on earthquake engineering, Keynote/Invited lecture, Lisbon

Bayat F, Kowsari M, Halldorsson (2024) A simplified seismicity model of the bookshelf fault system of the Southwest Iceland transform zone. Bulletin of Earthquake Engineering

Bayat F, Kowsari M, Halldorsson B (2022) A new 3-D finite-fault model of the Southwest Iceland bookshelf transform zone. Geophysical Journal International 231:1618–1633. https://doi.org/10.1093/gji/ggac272

Bayraktar B, Özer Sözdinler C, Necmioğlu Ö, Meral Özel N (2017) Preparation of Synthetic Earthquake Catalogue and Tsunami Hazard Curves in Marmara Sea using Monte Carlo Simulations. In: EGU General Assembly Conference Abstracts. p 1374

Bradley BA, Razafindrakoto H, Maurer B, et al (2018) Simulation-based ground motion prediction of historical and future New Zealand earthquakes and consequent geohazard impacts. In: Geotechnical Earthquake Engineering and Soil Dynamics V. American Society of Civil Engineers Reston, VA, pp 29–42

Callaghan S, Goulet C, Silva F, et al (2021) Verification and Validation of the Broadband CyberShake Platform. In: AGU Fall Meeting Abstracts. pp S11A-05

Callaghan S, Maechling P, Deelman E, et al (2008) Reducing time-to-solution using distributed high-throughput mega-workflows-experiences from SCEC CyberShake. In: 2008 IEEE Fourth International Conference on eScience. IEEE, pp 151–158

Callaghan S, Maechling PJ, Goulet CA, et al (2020) Enhancing CyberShake Simulations for Engineering Applications. 2020:S068-05

Callaghan S, Maechling PJ, Goulet CA, et al (2017) CyberShake Physics-Based PSHA in Central California. In: AGU Fall Meeting Abstracts. pp NH23B-04

Callaghan S, Maechling PJ, Juve G, et al (2015) Using CyberShake Workflows to Manage Big Seismic Hazard Data on Large-Scale Open-Science HPC Resources. In: AGU Fall Meeting Abstracts. pp IN43B-1738

Callaghan S, Maechling PJ, Juve G, et al (2014) Optimizing CyberShake Seismic Hazard Workflows for Large HPC Resources. In: AGU Fall Meeting Abstracts. pp IN21C-3720

Clifton AE, Pagli C, Jónsdóttir JF, et al (2003) Surface effects of triggered fault slip on Reykjanes Peninsula, SW Iceland. Tectonophysics 369:145–154

Cui Y, Olsen KB, Jordan TH, et al (2010) Scalable Earthquake Simulation on Petascale Supercomputers. In: SC '10: Proceedings of the 2010 ACM/IEEE International Conference for High Performance Computing, Networking, Storage and Analysis. pp 1–20

Darbyshire FA, White RS, Priestley KF (2000) Structure of the crust and uppermost mantle of Iceland from a combined seismic and gravity study. Earth and Planetary Science Letters 181:409–428

Decriem J, Árnadóttir T, Hooper A, et al (2010) The 2008 May 29 earthquake doublet in SW Iceland. Geophysical Journal International 181:1128–1146

Dubois L, Feigl KL, Komatitsch D, et al (2008) Three-dimensional mechanical models for the June 2000 earthquake sequence in the south Iceland seismic zone. Tectonophysics 457:12–29. https://doi.org/10.1016/j.tecto.2008.05.020

Einarsson P (1991) Earthquakes and present-day tectonism in Iceland. Tectonophysics 189:261–279

Einarsson P (2014) Mechanisms of Earthquakes in Iceland. In: Beer M, Kougioumtzoglou IA, Patelli E, Au IS-K (eds) Encyclopedia of Earthquake Engineering. Springer Berlin Heidelberg, Berlin, Heidelberg, pp 1–15

Einarsson P (2008) Plate boundaries, rifts and transforms in Iceland. Jökull 58:35–58

Einarsson P (2010) Mapping of Holocene surface ruptures in the South Iceland Seismic Zone. Jökull 60:121–138

Einarsson P (2015) Mechanisms of earthquakes in Iceland. Encyclopedia of Earthquake Engineering 1–15

Einarsson P, Björnsson S, Foulger G, et al (1981) Seismicity pattern in the South Iceland seismic zone. Earthquake Prediction 141–151

Einarsson P, Hjartardóttir ÁR, Hreinsdóttir S, Imsland P (2020) The structure of seismogenic strike-slip faults in the eastern part of the Reykjanes Peninsula Oblique Rift, SW Iceland. Journal of Volcanology and Geothermal Research 391:106372. https://doi.org/10.1016/j.jvolgeores.2018.04.029

Field EH, Dawson TE, Felzer KR, et al (2009) Uniform California earthquake rupture forecast, version 2 (UCERF 2). Bulletin of the Seismological Society of America 99:2053–2107

Graves R, Jordan TH, Callaghan S, et al (2011) CyberShake: A physics-based seismic hazard model for southern California. Pure and Applied Geophysics 168:367–381

Graves R, Pitarka A (2016) Kinematic ground-motion simulations on rough faults including effects of 3D stochastic velocity perturbations. Bulletin of the Seismological Society of America 106:2136–2153

Graves RW, Pitarka A (2010) Broadband ground-motion simulation using a hybrid approach. Bulletin of the Seismological Society of America 100:2095–2123

Halldórsson B, Ólafsson S, Sigbjörnsson R (2007) A fast and efficient simulation of the far-fault and near-fault earthquake ground motions associated with the June 17 and 21, 2000, earthquakes in South Iceland. Journal of Earthquake Engineering 11:343–370

Halldorsson B, Sigbjörnsson R (2009) The $M_w$6.3 Ölfus earthquake at 15:45 UTC on 29 May 2008 in South Iceland: ICEARRAY strong-motion recordings. Soil Dynamics and Earthquake Engineering 29:1073–1083. https://doi.org/10.1016/j.soildyn.2008.12.006

Hjaltadóttir S (2009) Use of relatively located microearthquakes to map fault patterns and estimate the thickness of the brittle crust in Southwest Iceland. Master's thesis, Faculty of Earth Sciences, University of Iceland, Reykjavík, Iceland

Hreinsdóttir S, Árnadóttir T, Decriem J, et al (2009) A complex earthquake sequence captured by the continuous GPS network in SW Iceland. Geophysical Research Letters 36:L12309

Jónasson K, Bessason B, Helgadóttir Á, et al (2021) A harmonised instrumental earthquake catalogue for Iceland and the northern Mid-Atlantic Ridge. Nat Hazards Earth Syst Sci 21:2197–2214. https://doi.org/10.5194/nhess-21-2197-2021

Jordan TH, Callaghan S (2018) CyberShake models of seismic hazards in Southern and Central California. In: Proceedings of the US National Conference on Earthquake Engineering

Karaca H (2021) Generation of synthetic catalog by using Markov chain Monte Carlo simulation and inverse Poisson distribution. Journal of Seismology 25:1103–1114

Kowsari M, Ghasemi S, Bayat F, Halldorsson B (2022a) A backbone seismic ground motion model for strike-slip earthquakes in Southwest Iceland and its implications for near-and far-field PSHA. Bulletin of Earthquake Engineering 1–24. https://doi.org/10.1007/s10518-022-01556-z

Kowsari M, Halldorsson B, Bayat F (2022b) Preparation of Finite-fault Earthquake Catalogues Enabling Physics-based PSHA in Southwest Iceland. In: Proceedings of the 3rd European Conference on Earthquake and Engineering Seismology (3ECEES). Bucharest, Romania, pp 3863-3870 (No. 8441)

Kowsari M, Halldorsson B, Snæbjörnsson JÞ, Jónsson S (2021) Effects of different empirical ground motion models on seismic hazard maps for North Iceland. Soil Dynamics and Earthquake Engineering 148:106513. https://doi.org/10.1016/j.soildyn.2020.106513

Kowsari M, Sonnemann T, Halldorsson B, et al (2020) Bayesian Inference of Empirical Ground Motion Models to Pseudo-Spectral Accelerations of South Iceland Seismic Zone Earthquakes based on Informative Priors. Soil Dynamics and Earthquake Engineering 132:106075. https://doi.org/10.1016/j.soildyn.2020.106075

Li B, Gabriel A-A, Ulrich T, et al (2023) Dynamic rupture models, fault interaction and ground motion simulations for the segmented Húsavík-Flatey Fault Zone, Northern Iceland. Journal of Geophysical Research: Solid Earth 128:e2022JB025886. https://doi.org/10.1029/2022JB025886

Maechling P, Deelman E, Zhao L, et al (2007) SCEC CyberShake Workflows—Automating Probabilistic Seismic Hazard Analysis Calculations. In: Taylor IJ, Deelman E, Gannon DB, Shields M (eds) Workflows for e-Science: Scientific Workflows for Grids. Springer, London, pp 143–163

Mai PM, Beroza GC (2002) A spatial random field model to characterize complexity in earthquake slip. Journal of Geophysical Research: Solid Earth 107:ESE-10

Mai PM, Beroza GC (2000) Source scaling properties from finite-fault-rupture models. Bulletin of the Seismological Society of America 90:604–615

Mavroeidis GP, Dong G, Papageorgiou AS (2004) Near-fault ground motions, and the response of elastic and inelastic single-degree-of-freedom(SDOF) systems. Earthquake Engng Struct Dyn 33:1023–1049. https://doi.org/10.1002/eqe.391

Mavroeidis GP, Papageorgiou AS (2003) A Mathematical Representation of Near-Fault Ground Motions. Bulletin of the Seismological Society of America 93:1099–1131. https://doi.org/10.1785/0120020100

Milner K, Shaw B, Goulet C, et al (2021) Toward Physics-Based Nonergodic PSHA: A Prototype Fully Deterministic Seismic Hazard Model for Southern California. Bulletin of the Seismological Society of America 111:. https://doi.org/10.1785/0120200216

Milner KR, Callaghan S, Maechling PJ, et al (2018) A SCEC CyberShake Physics-Based Probabilistic Seismic Hazard Model for Northern California. In: AGU Fall Meeting Abstracts. pp S43D-0642

Morgan JP, Kleinrock MC (1991) Transform zone migration: Implications of bookshelf faulting at oceanic and Icelandic propagating ridges. Tectonics 10:920–935. https://doi.org/10.1029/90TC02481

O'Reilly O, Yeh T-Y, Olsen KB, et al (2022) A High-Order Finite-Difference Method on Staggered Curvilinear Grids for Seismic Wave Propagation Applications with Topography. Bulletin of the Seismological Society of America 112:3–22

Pagli C, Pedersen R, Sigmundsson F, Feigl KL (2003) Triggered fault slip on June 17, 2000 on the Reykjanes Peninsula, SWIceland captured by radar interferometry. Geophysical Research Letters 30:1273

Panzera F, Zechar JD, Vogfjörd KS, Eberhard DAJ (2016) A Revised Earthquake Catalogue for South Iceland. Pure Appl Geophys 173:97–116. https://doi.org/10.1007/s00024-015-1115-9

Pedersen R, Jónsson S, Árnadóttir T, et al (2003) Fault slip distribution of two June 2000 MW6.5 earthquakes in South Iceland estimated from joint inversion of InSAR and GPS measurements. Earth and Planetary Science Letters 213:487–502. https://doi.org/10.1016/S0012-821X(03)00302-9

Rodriguez JE, Monterrubio-Velasco M, Otilio R, d. l. P. J (2021) Cybershake migration and usage report

Roth F (2004) Stress Changes Modelled for the Sequence of Strong Earthquakes in the South Iceland Seismic Zone Since 1706. In: Fernández J (ed) Geodetic and Geophysical Effects Associated with Seismic and Volcanic Hazards. Birkhäuser, Basel, pp 1305–1327

Rupakhety R, Sigbjörnsson R (2013) Rotation-invariant measures of earthquake response spectra. Bulletin of Earthquake Engineering 11:1885–1893

Sigbjörnsson R, Baldvinsson GI, Thrainsson H (1995) A stochastic simulation approach for assessment of seismic hazard maps in "European Seismic Design Practice." Balkema, Rotterdam

Sigbjörnsson R, Ólafsson S, Rupakhety R, et al (2014) Strong-motion Monitoring and Accelerometric Recordings in Iceland. In: 2nd European Conference on Earthquake and Engineering Seismology (2ECEES). Istanbul, Turkey, 24-29 August, 2014, p Paper No. 2034

Silva F, Callaghan S, Maechling PJ, et al (2016) Expanding CyberShake Physics-Based Seismic Hazard Calculations to Central California. 2016:S31C-2778

Solnes J, Sigbjörnsson R, Eliasson J (2004) Probabilistic seismic hazard mapping of Iceland. In: 13th World conference on earthquake engineering (13WCEE). Vancouver, BC, Canada, August 1-6, 2004, p Paper No. 2337

Stefánsson R, Böðvarsson R, Slunga R, et al (1993) Earthquake prediction research in the South Iceland seismic zone and the SIL project. Bulletin of the Seismological Society of America 83:696–716

Stefánsson R, Bonafede M, Roth F, et al (2006) Modelling and parameterizing the Southwest Iceland earthquake release and deformation process. Icelandic Meteorological Office, Report No. VÍ-ES-03-06005

Stefansson R, Gudmundsson GB, Halldorsson P (2008) Tjörnes fracture zone. New and old seismic evidences for the link between the North Iceland rift zone and the Mid-Atlantic ridge. Tectonophysics 447:117–126

Steigerwald L, Einarsson P, Hjartardóttir ÁR (2020) Fault kinematics at the Hengill Triple Junction, SW-Iceland, derived from surface fracture pattern. Journal of Volcanology and Geothermal Research 391:106439

Thingbaijam KKS, Martin Mai P, Goda K (2017) New empirical earthquake source-scaling laws. Bulletin of the Seismological Society of America 107:2225–2246

Thordarson T, Hoskuldsson A (2002) Iceland: Classic Geology in Europe. Terra, United Kingdom

Tryggvason E, Thoroddsen S, Thorarinsson S (1958) Report on earthquake risk in Iceland. Timarit Verkfraedingafelags Islands 43:81–97

Tryggvason K, Husebye ES, Stefánsson R (1983) Seismic image of the hypothesized Icelandic hot spot. Tectonophysics 100:97–118. https://doi.org/10.1016/0040-1951(83)90180-4

Wells DL, Coppersmith KJ (1994) New empirical relationships among magnitude, rupture length, rupture width, rupture area, and surface displacement. Bulletin of the seismological Society of America 84:974–1002

23